\documentclass[12pt]{article}
\usepackage{feynmp}
\usepackage{array}
\usepackage{graphicx}
\newcommand{\indi}[1]{\textit{\scriptsize{#1}}}
\makeatletter
\def\fmslash{\@ifnextchar[{\fmsl@sh}{\fmsl@sh[0mu]}}
\def\fmsl@sh[#1]#2{%
  \mathchoice
    {\@fmsl@sh\displaystyle{#1}{#2}}%
    {\@fmsl@sh\textstyle{#1}{#2}}%
    {\@fmsl@sh\scriptstyle{#1}{#2}}%
    {\@fmsl@sh\scriptscriptstyle{#1}{#2}}}
\def\@fmsl@sh#1#2#3{\m@th\ooalign{$\hfil#1\mkern#2/\hfil$\crcr$#1#3$}}
\makeatother
\global\arraycolsep=2pt 
\input{epsf}
\usepackage{bbm}
\usepackage{latexsym}
\begin{document}
\thispagestyle{empty}
\rightline{TTP03-15}
\rightline{hep-ph/0306043}
\rightline{\today}
\bigskip
\boldmath
\begin{center}
{\bf \Large
Towards a Generic Parametrisation of \\[2mm] ``New Physics'' 
in Quark-Flavour Mixing}
\end{center}
\unboldmath
\smallskip
\begin{center}
{\large{\sc Thomas Hansmann,  Thomas Mannel}}
\vspace*{2cm} \\ 
{\sl Institut f\"{u}r Theoretische Teilchenphysik, \\
Universit\"{a}t Karlsruhe,  D--76128 Karlsruhe, Germany}
\end{center}
\vfill
\begin{abstract}
\noindent
We consider a special class of dimension-six operators which are
assumed to be induced by some new physics with typical scales of
order $\Lambda$. This special class are the operators with two quarks
which can mediate transitions between quark flavours. We show that under
quite general assumptions the effect of these operators can be parametrised
in terms of six parameters, leading to a modification of the (at tree level
flavour diagonal) neutral currents and to an ``effective CKM matrix''
for the charged currents, which is
not necessarily unitary any more. The effects of these operators
on charged and neutral currents are studied. 
\end{abstract}
\newpage
\section{Introduction}
In the coming few years the flavour sector of the standard model (SM)
will have to pass its first detailed test, which hopefully leads to some
hint to physics beyond the SM. Being the most general renormalizable
theory compatible with the observed broken
$SU(3)_C \times SU(2)_L \times U(1)_Y$ symmetry and the observed particle
spectrum, any effect beyond the SM has to show up at the scale of the weak
boson masses as a set of operators with mass dimension of six or higher,
where these operators have to be compatible with the SM symmetry.
The coupling constants of the dimension-six operators scale as
$1/\Lambda^2$ where $\Lambda$ represents the scale of the new physics.

All the possible operators have been listed already some time ago
\cite{BW}, but due
to their large number this approach is - in its full generality -
useless for phenomenological applications,  
since every operator comes with an unknown coupling constant. Thus it
is unavoidable to restrict their number by some assumption. As an
example one can consider the parametrisation of new-physics effects
in terms of the Peskin-Takeuchi parameters $S$, $T$ and
$U$ \cite{peskint}\footnote{Alternatively
                            $\epsilon_1$, $\epsilon_2$ $\epsilon_3$
                            have been used, which are closely related
			    to $S$, $T$ and $U$.}
which have been used in the analysis of the LEP precision data.
These parameters can be related to a certain subset of dimension-six
operators \cite{Georgi} and are thus an example for a generic analysis.

However, up to now flavour physics lacks such a simple parametrisation
of new-physics effects.
Looking at the list of dimension-six operators which can appear at
the scale of the weak bosons only those with quark
fields\footnote{We do not consider the flavour physics in the leptonic sector
here, an extension to include this is obvious.}
are relevant for flavour physics. These operators have either four quark
fields (in which case there are no other fields) or two quark fields, in
which case the remaining three mass dimensions are made up by either
covariant derivatives or Higgs fields.

A specific feature of the SM is the significant suppression of neutral
currents by the GIM mechanism, making neutral current transitions very
sensitive to possible new-physics effects. This has been investigated before
in a number of publications; see eg. \cite{NCpapers,MFV}.
However, the symmetries of
the SM suggest that one could also have effects in the charged currents,
where the SM effects are at best CKM suppressed and thus one has less
sensitivity to new-physics effects. 

In the present paper we discuss the impact of dimension-six two-quark
operators, which could be induced at the scale of the weak bosons
by some new-physics effect. We are aiming at a simple phenomenological
parametrisation in the spirit of the aforementioned analysis of the
gauge sector by Peskin and Takeuchi. A similar aaproach has been suggested
recently for the Higgs sector \cite{Barger}.

In the next section we classify the general dimension-six operators
which are relevant at the scale of the weak boson mass, which are
bilinear in the quark fields and which are compatible with the
symmetries of the SM. We make well defined simplifying assumptions
that restrict the number of parameters to only 6. Finally we discuss
our result and conclude.

\section{Dimension-Six Operators}
We shall first write down the standard model contributions in order
to fix our notation. 
Starting from a $SU(2)_L \times SU(2)_R$ symmetry we group the
left-handed quarks according to
\begin{equation}
Q_1 = \left(\begin{array}{c} u_L \\ d_L \end{array} \right) \quad
Q_2 = \left(\begin{array}{c} c_L \\ s_L \end{array} \right) \quad
Q_3 = \left(\begin{array}{c} t_L \\ b_L \end{array} \right)
\end{equation}
and likewise for the right-handed quarks
\begin{equation}
q_1 = \left(\begin{array}{c} u_R \\ d_R \end{array} \right) \quad
q_2 = \left(\begin{array}{c} c_R \\ s_R \end{array} \right) \quad
q_3 = \left(\begin{array}{c} t_R \\ b_R \end{array} \right)
\end{equation}
such that $Q_A$ transforms as a $(2,1)$ and $q_A$ as a $(1,2)$ under
$SU(2)_L \times SU(2)_R$.

The Higgs field transforms as a $(2,2)$ under this symmetry and we gather
the two real fields $\phi_0$, $\chi_0$ and the complex field
$\phi_+ = \phi_-^*$  into a  $ 2 \times 2$ matrix\footnote{
     Here we assume a linear representation of the electroweak symmetry;
     however, we express everything in terms of the field $H$ in terms of
     which one can easily switch to the non-linear representation by the
     replacement $H \to (v / \sqrt{2}) \Sigma$ where $\Sigma$ is the
     (matrix valued) field of the non-linear sigma model.}
\begin{equation}
H = \frac{1}{\sqrt{2}}
\left( \begin{array}{cc}
           \phi_0 - i \chi_0   &\   \sqrt{2} \phi_+ \\
           - \sqrt{2} \phi_-   &\   \phi_0 + i \chi_0 \end{array} \right)\,.
\end{equation}

We can now write down the Lagrangian for a non-linear sigma model,
having an $SU(2)_L \times SU(2)_R$ symmetry, broken down to the diagonal
$SU(2)_{L+R}$ by vacuum-expectation value $v$ of the Higgs field
$$
\langle 0 | H | 0 \rangle = \frac{v}{\sqrt{2}} \mathbbm{1}
$$
leading to mass terms for the quarks. 
This corresponds almost to the (ungauged) standard model,
except that $SU(2)_L \times SU(2)_R$ is explicitly broken down to 
$SU(2)_L \times U(1)_Y$ by the mass terms of the quarks. Note that 
the Higgs sector itself still has the full $SU(2)_L \times SU(2)_R
\to SU(2)_{L+R} \equiv SU(2)_C$ symmetry, which is the well known custodial
symmetry. 
%
The renormalisable Lagrangian invariant under $SU(2)_L \times U(1)_Y$
corresponding to the ungauged standard model is 
\begin{eqnarray} \label{SM}
{\cal L}_{SM} &=& \overline{Q}_A (i\fmslash{\partial}) Q_A
                + \overline{q}_A (i\fmslash{\partial}) q_A
                - \frac{1}{v}(\overline{Q}_A H \hat{M}_{AB} q_B + h.c.)
\nonumber \\
    &+& \frac{1}{2} {\rm Tr} \left\{ (\partial_\mu H)^\dagger
                                     (\partial^\mu H) \right\}
     -  V \left(  {\rm Tr} \left\{H^\dagger H \right\} \right)
\end{eqnarray}
where we defined the mass matrix 
\begin{equation} \label{MassMat}
\hat{M}_{AB} = \frac{1}{2} (m^u_{AB} + m^d_{AB}) \mathbbm{1} +
                       \frac{1}{2} (m^u_{AB} - m^d_{AB}) \tau^3
\end{equation}
where $m^{u/d}$ correspond to the mass matrices of the up/down-type quarks. 
Note that the term proportional to $\tau^3$ explicitly breaks
$SU(2)_C$, leading to a splitting between up- and down-quark masses
and to mixing between families.

The standard model is obtained from gauging the 
$SU(2)_L \times U(1)_Y$ symmetry, which means that the ordinary
derivatives have to be replaced by covariant ones
\begin{eqnarray}
i D_\mu Q_A &=& i \partial_\mu Q_A
                + \frac{1}{\sqrt{2}} g \left( \tau^+  W_\mu^+
                  + \tau^-  W_\mu^- \right)\, Q_A
                + \frac{1}{2} g \tau^3 W_\mu^3 Q_A \nonumber  \\
&& \qquad \quad                 + \frac{1}{6} g' B_\mu\, Q_A \\
i D_\mu q_A &=& i \partial_\mu q_A + \frac{1}{2} g'
                \left(\frac{1}{3} + \tau^3 \right) B_\mu\, q_A \\
i D_\mu H &=& i \partial_\mu H + \frac{1}{2} g \tau^a W_\mu^a H
                                 - \frac{1}{2} g' B_\mu H \tau^3
\end{eqnarray}
The physical fields for the neutral bosons are obtained
by the usual rotation $W_\mu^3 = \cos \Theta_W Z_\mu + \sin \Theta_W
A_\mu$, $B_\mu = \cos \Theta_W A_\mu - \sin \Theta_W Z_\mu$ and $g \sin
\Theta_W = g' \cos \Theta_W$.

Eq. (\ref{SM}) (more precisely its gauged version)
is the most general renormalizable Lagrangian with an
$SU(2)_L \times U(1)_Y$ symmetry and with this particle content.
Going beyond the standard model means to consider operators of dimension 
higher than four. It turns out that there are no dimension-five
operators compatible with $SU(2) \times U(1)$ symmetry. The 
dimension-six operators appear 
with couplings suppressed by two powers of the scale of new physics $\Lambda$. 
We are going to consider those dimension-six operators which involve two
quark fields. They may be classified according to the helicities of the
quark fields: left-left (LL), right-right (RR)
and left-right (LR). Therefore the Lagrangian is given by 
\begin{equation} \label{GeneralL}
  \mathcal{L} = \mathcal{L}_{SM} 
+ \frac{1}{\Lambda^2} \sum_i O^{(i)}_{LL}
+ \frac{1}{\Lambda^2} \sum_i O^{(i)}_{RR}
+ \frac{1}{\Lambda^2} \sum_i O^{(i)}_{LR}
\end{equation}
The two quark fields have in total dimension three,
the remaining dimensionality has to come either from covariant derivatives
or from powers of the Higgs field. Note that (\ref{GeneralL}) is in fact very
general; possible exceptions are e.g.\ special models with more than one Higgs
doublet. 

Furthermore, we do not include QCD into our discussion, 
since we assume that strong interactions to be flavour blind.
A generalisation of the
operator basis would be straightforward.
Thus we have the following operators \cite{BW,espriu}: 
\paragraph{LL-Operators with three derivatives:}
\begin{eqnarray}
O^{(1)}_{LL} &=& G_{AB}^{(1)}\, \overline{Q}_A \left(i \fmslash{D}\right)^3 Q_B \\
O^{(2)}_{LL} &=& G_{AB}^{(2)} \, \overline{Q}_A
         \left\{ i \fmslash{D} \, , \,\sigma^{\mu\nu}B_{\mu\nu}
	 \right\} Q_B \\
O^{(3)}_{LL} &=& i G_{AB}^{(3)} \, \overline{Q}_A
         \left[ i \fmslash{D} \, , \, \sigma^{\mu\nu}B_{\mu\nu} \right] Q_B \\
O^{(4)}_{LL} &=& G_{AB}^{(4)} \, \overline{Q}_A 
         \left\{ i \fmslash{D} \, , \, \sigma^{\mu\nu}W_{\mu\nu} \right\} Q_B \\
O^{(5)}_{LL} &=& i G_{AB}^{(5)} \, \overline{Q}_A
         \left[ i \fmslash{D} \, , \, \sigma^{\mu\nu}W_{\mu\nu} \right] Q_B \\
O^{(6)}_{LL} &=& G_{AB}^{(6)} \overline{Q}_A \left[ i D^\mu \, , \,
  i B_{\mu\nu} \right] \gamma^\nu Q_B\\ 
O^{(7)}_{LL} &=& G_{AB}^{(7)} \overline{Q}_A \left[ i D^\mu \, , \,
  i W_{\mu\nu} \right] \gamma^\nu Q_B
\end{eqnarray}
Where the matrices $G^{(i)}$ are hermitean and $B_{\mu\nu}$ and
$W_{\mu\nu}$ are the field strength tensors of the $U(1)_Y$ and the
$SU(2)_L$ symmetries. Since field strength tensors can be written as commutators of
covariant derivatives they are counted as two derivatives. For
product groups (like in the SM) the commutator gives only
a linear combination of the field strength tensors, so we have to
treat them separately avoiding more than one covariant derivative per
operator. Note that the covariant derivative acts on all fields
to the right.

\paragraph{LL-Operators with two Higgs fields
           and one derivative:}\mbox{}\\
For these operators we get
\begin{eqnarray}
O^{(8)}_{LL} &=& \overline{Q}_A
         \left\{ H \hat{G}_{AB}^{(8)} H^\dagger \, , \,
                 (i \fmslash{D}) \right\} Q_B \\
O^{(9)}_{LL} &=& i \overline{Q}_A
         \left[ H \hat{G}_{AB}^{(9)} H^\dagger \, , \,
                 (i \fmslash{D}) \right] Q_B \\
O^{(10)}_{LL} &=&  \overline{Q}_A
          H \hat{G}_{AB}^{(10)} (i \fmslash{D}) H^\dagger Q_B
\end{eqnarray}
The matrices $\hat{G}_{AB}^{(i)}$ are hermitean and consist of a $SU(2)_C$
conserving and a $SU(2)_C$ breaking piece as
\begin{equation}
\hat{G}_{AB}^{(i)} = G_{AB}^{(i)}  \mathbbm{1} + G_{AB}^{(i) \prime} \tau_3.
\end{equation}

\paragraph{RR-Operators with three derivatives:}\mbox{}\\
In this case we have more operators due to the possibility of
explicitly breaking custodial $SU(2)_C$. These additional operators
are obtained by replacing all occurrences of $q_A$ by $\tau^3 q_A$
in all possible ways. Thus we shall only write the $SU(2)_C$-conserving
operators, which are
\begin{eqnarray}
O^{(1)}_{RR} &=& F_{AB}^{(1)} \, \overline{q}_A \left(i \fmslash{D}\right)^3 q_B
\\
O^{(2)}_{RR} &=& F_{AB}^{(2)} \, \overline{q}_A
         \left\{ i \fmslash{D} \, , \, \sigma^{\mu\nu}B_{\mu\nu} \right\} q_B \\
O^{(3)}_{RR} &=& i F_{AB}^{(3)} \, \overline{q}_A
         \left[ i \fmslash{D} \, , \, \sigma^{\mu\nu}B_{\mu\nu} \right] q_B \\
O^{(4)}_{RR} &=& F_{AB}^{(4)} \, \overline{q}_A \left[ i D^\mu \, , \,
  i B_{\mu\nu} \right]\gamma^\nu  q_B
\end{eqnarray}

\paragraph{RR-Operators with two Higgs fields
           and one derivative:}\mbox{}\\
In the same way we get for the custodial $SU(2)_C$ conserving operators
\begin{eqnarray}
O^{(5)}_{RR} &=& F_{AB}^{(5)} \, 
     \overline{q}_A\,
\left\{ H^\dagger H \, , \, (i \fmslash{D}) \right\} \,q_B \\
O^{(6)}_{RR} &=& iF_{AB}^{(6)} \,
     \overline{q}_A\,
\left[ H^\dagger H \, , \, (i \fmslash{D}) \right] \,q_B \\
O^{(7)}_{RR} &=&  F_{AB}^{(7)} \, 
     \overline{q}_A\, \,H^\dagger (i \fmslash{D}) H \, q_B
\end{eqnarray}
Again all the matrices $F^{(i)}$ have to be hermitean and the custodial
$SU(2)_C$ violating operators are obtained by the replacement
$q_A \longrightarrow \tau^3 q_A$

\paragraph{LR-Operators with three Higgs fields:}\mbox{}\\
Here we have only one operator conserving custodial $SU(2)_C$
\begin{eqnarray}\label{threeH}
O^{(1)}_{LR} &=&  K_{AB}^{(1)} \overline{Q}_A H H^\dagger H q_B + h.c. 
\end{eqnarray}
Note that the matrix $K^{(1)}$ needs not to be hermitean. There is also
one custodial $SU(2)_C$ violating operator which is obtained by the replacement
$q_A \longrightarrow \tau^3 q_A$.

\paragraph{LR-Operators with one Higgs field
           and two derivatives:}
\begin{eqnarray}
O^{(2)}_{LR} &=& \overline{Q}_A\, H \hat{K}_{AB}^{(2)}
            (i\fmslash{D})^2 \,q_B + h.c. \\
O^{(3)}_{LR} &=& \overline{Q}_A\, (i\fmslash{D})^2
            H \hat{K}_{AB}^{(3)}\,q_B + h.c. \\
O^{(4)}_{LR} &=& \overline{Q}_A\, \sigma^{\mu\nu}B_{\mu\nu} H \hat{K}_{AB}^{(4)} \,q_B + h.c. \\
O^{(5)}_{LR} &=& \overline{Q}_A\, \sigma^{\mu\nu}W_{\mu\nu} H \hat{K}_{AB}^{(5)} \,q_B + h.c. \\
O^{(6)}_{LR} &=& \overline{Q}_A\,
        (i\fmslash{D}) H \hat{K}_{AB}^{(6)} (i\fmslash{D}) \,q_B + h.c. \\
O^{(7)}_{LR} &=& \overline{Q}_A\, (iD_\mu)
            H \hat{K}_{AB}^{(7)}(iD^\mu) \,q_B + h.c.
\end{eqnarray}
where we again use the notation 
\begin{equation}
\hat{K}_{AB}^{(i)} = K_{AB}^{(i)}  \mathbbm{1} + K_{AB}^{(i) \prime} \tau_3.
\end{equation}
to include the custodial $SU(2)_C$ violating contributions. 
As in (\ref{threeH}) the matrices $K^{(i)}_{AB}$ and
$K^{(i)\prime}_{AB}$ need not to be hermitean.

We shall not discuss the renormalisation of these operators. In fact, the
set of operators we have listed does not close under renormalisation. This is
obvious, since we leave out the four-quark operators, some of which mix into
the operators listed above and vice versa. However, we do not consider this to
a be a problem, since renormalisation effects are small due to small couplings.

\section{Reducing the number of operators}
In this section we will discuss under which assumptions the number 
of operators can be reduced. We try to keep these
assumptions as general as possible in order to obtain a generic
parametrisation of possible new-physics effects.

\subsection{Equations of Motion}
The operators listed above are not independent, since some of them
are connected by the equations of motion (eom)
\begin{eqnarray} \label{eom1}
(i\fmslash{D}) Q_A &=& \frac{1}{v} H {\cal M}_{AB} \, q_B \\\label{eom2}
(i\fmslash{D}) q_A &=& \frac{1}{v} {\left(\cal M^\dag\right)}_{AB}
                       H^\dagger \, Q_B\,.
\end{eqnarray}
This allows us to eliminate all the operators with two quarks and
three covariant derivatives, with the exception of $O^{(6)}_{LL}$,
$O^{(7)}_{LL}$ and
$O^{(4)}_{RR}$. These can be rewritten as four-fermion operators by the
equation of motion for the gauge fields, and hence they are not in
the class of operators we are considering here. 

The same is true for dimension-six operators involving the gluonic
field strength; we have not explicitly written these operators but,  
since the gluon momenta are again of the order of the
masses of the external states, they can be dropped for the same reason.

Using the equations of motion allows us in particular to remove all
operators which yield (after spontaneous symmetry breaking) a contribution
to the (irreducible) two-point vertex functions corresponding to the
kinetic energy. This is a natural choice, since these contributions
correspond only to field redefinitions, mass renormalisations and
to a renormalisation of the CKM matrix.

This leads to the following basis of operators:
\subparagraph{LL-Operators}
\begin{eqnarray} \label{LL1}
O^{(1)}_{LL} &=& \overline{Q}_A\, \fmslash{L} G_{AB}^{(1)} \,Q_B \\ \label{LL2}
O^{(2)}_{LL} &=& \overline{Q}_A\, \fmslash{L_3}G_{AB}^{(2)} \,Q_B 
\end{eqnarray}
with
\begin{eqnarray}
L^\mu &=& H \left(i D^\mu H\right)^\dag +  \left(i D^\mu H\right)  H^\dag \\
L^\mu_3 &=& H \tau_3 \left(i D^\mu H\right)^\dag +
            \left(i D^\mu H\right)\tau_3   H^\dag 
\end{eqnarray}
and all matrices $G^{(i)}_{AB}$ being hermitean.

\subparagraph{RR-Operators}
\begin{eqnarray}
O^{(1)}_{RR} &=& \overline{q}_A\, \fmslash{R}F_{AB}^{(1)} \,q_B \\
O^{(2)}_{RR} &=& \overline{q}_A\, \left\{ \tau_3,
            \fmslash{R}\right\} F_{AB}^{(2)} \,q_B \\ 
O^{(3)}_{RR} &=& i\overline{q}_A\, \left[  \tau_3,
        \fmslash{R}\right]  F_{AB}^{(3)}\,q_B \\
O^{(4)}_{RR} &=& \overline{q}_A\, \tau_3 \fmslash{R} \tau_3 F_{AB}^{(4)} \,q_B
\end{eqnarray}
with
\begin{eqnarray}
R^\mu &=& H^\dag \left(i D^\mu H\right)+  \left(i D^\mu H\right)^\dag  H
\end{eqnarray}
and again hermitean $F_{AB}^{(i)}$.

\subparagraph{LR-Operators}
\begin{eqnarray}
O^{(1)}_{LR} &=& \overline{Q}_A\, HH^\dag H \hat{K}_{AB}^{(1)} \,q_B + h.c.
\label{hhh} \\
O^{(2)}_{LR} &=& \overline{Q}_A\, \left(\sigma_{\mu\nu}B^{\mu \nu}  \right)H
\hat{K}_{AB}^{(2)} \,q_B \,  + h.c.\\
O^{(3)}_{LR} &=& \overline{Q}_A\,
\left(\sigma_{\mu\nu} W^{\mu\nu} \right) H\hat{K}_{AB}^{(3)} \,q_B + h.c.\\
O^{(4)}_{LR} &=& \overline{Q}_A\,
\left(iD_\mu H\right) iD^\mu \hat{K}_{AB}^{(4)} \,q_B + h.c.
\end{eqnarray}
The coupling matrices  
\begin{equation}
\hat{K}_{AB}^{(i)} = {K}^{(i)}_{AB} + \tau_3 {K}_{AB}^{(i)\prime}
\end{equation}
do not have to be hermitean.

\subsection{Chiral limit}
In the standard model any coupling between left-handed and
right-handed  components of the fields is only due to the mass term. This
implies that (except for the top quark) the corresponding Yukawa
couplings are very small.

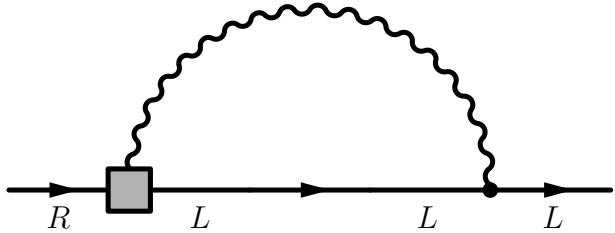
\begin{figure}
  \begin{center}
    \begin{fmffile}{HMmass}
\begin{fmfgraph*}(100,30)
  \fmfpen{thick}
  \fmfleft{i1,o1}
  \fmfright{i2,o2}
  \fmf{fermion, label=$L$,label.side=right}{v2,i2}
  \fmf{plain,label=$L$}{v3,v2}
  \fmf{fermion}{v4,v3}
  \fmf{plain,label=$L$}{v5,v4}
  \fmf{fermion,label=$R$}{i1,v5}
  \fmfdot{v2}
  \fmfv{decor.shape=square,decor.size=8thick,decor.filled=30}{v5}
  \fmffreeze
  \fmf{boson,left}{v5,v2}
  \end{fmfgraph*}
\end{fmffile}\\[2mm]
\caption{Diagram which induces a mass term from e.g. ${\mathcal O}^{(3)}_{LR}$.
         The shaded box denotes an insertion of ${\mathcal O}^{(3)}_{LR}$}
\label{fig1}
\end{center}
\end{figure}

Any dimension-six operator coupling left and right helicities will
(after spontaneous symmetry breaking) either contain a mass term
(such as $O^{(1)}_{LR}$) or it will mix into a mass term. In fact, evaluating
diagrams of the type shown in fig.~\ref{fig1} will lead to a quadratic
divergence which means that these operators will mix with the
dimension-four mass term of the original Lagrangian (\ref{SM}).
In particular, even if the mass term in (\ref{SM}) were absent, the
operators $O^{(i)}_{LR}$ would induce a mass of the order
$$
m^{(i)}_{LR} \sim \frac{g^2}{16 \pi^2} v \, g^{(i)}_{LR}
$$
where $g$ is the coupling of $SU(2)_L \times U(1)_Y$ and 
$g^{(i)}_{LR}$ are the couplings of the $O^{(i)}_{LR}$, i.e. one of the
matrix elements of $\hat K^{(i)}_{AB}$. In order to
comply with the observed smallness of the Yukawa couplings we are
led to our first assumption: {\em We assume that a chiral limit exists, in
which all quarks become massless, even if the dimension-six
contributions
are included.}

Formally we can implement this limit by replacing each occurrence of the
combination $H q_A$ or $H \tau^3 q_A$ by
\begin{equation}
{ H q_A, \quad H \tau^3 q_A } \quad \longrightarrow \quad 
{ \lambda H q_A, \quad \lambda H \tau^3 q_A }
\end{equation}
where $\lambda$ is a parameter which vanishes in the chiral limit.
This replacement happens also in
in the dimension-four piece  (\ref{SM}) corresponding to the
standard model; if the Yukawa couplings in (\ref{SM})
were of order unity, we could chose $\lambda$ to be of order
$m_q / v$ to achive the smallness of the Yukawa couplings of the (light)
quarks. Likewise, this parameter will also multiply the $O^{(i)}_{LR}$,
making these couplings small as well. 

Thus we argue that in order to keep the light quark masses small
the natural assumption for the couplings $\hat K^{(i)}$ is
\begin{equation}
g^{(i)}_{LR} \propto K^{(i)} \sim \lambda = \frac{m_q}{v} 
\end{equation}
which will make the contributions of all $O^{(i)}_{LR}$ very small. 

However, the SM contribution to the neutral currents mediated by
$O^{(i)}_{LR}$ is also suppressed by the GIM mechanism as well as
by loop factors, so for the neutral currents the net-supression factor
of the SM contributions relative to the the new-physics contribution is
$v^2 / \Lambda^2$. Still the absolute size of the effects is small
and we shall neglect these contributions in the following.

We may use the above argument (although here it may be weaker) to
consider the $O^{(i)}_{RR}$ operators. If new-physics contributions are
purely left-handed (i.e. if we have only new interactions acting on the
left-handed components), this would imply that all the $O^{(i)}_{RR}$ are
suppressed by two powers of $\lambda$.
\begin{equation}
F^{(i)} \sim \lambda^2 = \left(\frac{m_q}{v}\right)^2
\end{equation}
Although this may be considered very restrictive  
we still will make this assumption and neglect also the operators
of the type $O^{(i)}_{RR}$.

Using this assumption we continue with assuming that the
operators $O_{LL}^{(1)}$ and $O_{LL}^{(2)}$ shown in (\ref{LL1}) and (\ref{LL2}) yield the leading
new-physics contribution. In particular, after spontaneous symmetry
breaking one finds an anomalous quark-gauge boson coupling
of the order $v^2 / \Lambda^2$.

\subsection{Flavour Conservation}
The next  assumption  we are going to discuss is flavour conservation.
It is well known that flavour-changing neutral currents (FCNCs)
are very strongly suppressed. In the standard model the violation
of flavour occurs through the fact that
\begin{equation}
\left[ m^u m^{u\dag} \, , \, m^d m^{d\dag} \right] \neq 0 
\end{equation}
implying that there is no basis in the left-handed flavour
space where both mass matrices
are diagonal. FCNCs are suppressed in the SM by the GIM mechanism
\cite{GIM}. It ensures that only the mass differences between up- or
down-type quarks are relevant for FCNCs. Therefore significant contributions
in the SM can come from the top quark only, but these suffer from
an additional suppression by the small mixing angles.

For the discussion of the dimension-six operators we shall use a concept
similar to minimal flavour violation \cite{MFV}, which means that all the FCNCs
induced by tree level contributions of the dimension-six operators
are assumed to vanish, i.e. the corresponding coupling matrices are
diagonal in the mass eigenbasis.

In particular, for our leading contribution shown in (\ref{LL1}) and
(\ref{LL2}) we find after spontaneous symmetry breaking
\begin{eqnarray} \label{additional}
&&H (i D_\mu H)^\dagger +  (i D_\mu H)  H^\dagger = \\\nonumber
&&\qquad\qquad v^2 \frac{g}{\sqrt{2}}
\left(\tau^+ W_\mu^+ + \tau^- W_\mu^- \right) 
+ v^2 \frac{g}{2 \cos \theta_W} Z_\mu \tau^3 
+ \cdots  \\
&&H \tau^3 (i D_\mu H)^\dagger
              +  (i D_\mu H)  \tau^3 H^\dagger 
=  v^2 \frac{g}{2 \cos \theta_W} Z_\mu + \cdots.
\end{eqnarray}
The neutral currents from the operators in (\ref{LL1}) and
(\ref{LL2}) yield
\begin{equation} \label{nc}
O^{(1)}_{LL} + O^{(2)}_{LL} = v^2 \frac{g}{2 \cos \theta_W} Z_\mu
\overline{Q}_A \gamma^\mu \tau^3
\left(G_{AB}^{(1)} + \tau^3 G_{AB}^{(2)} \right) Q_B + \cdots.
\end{equation}
where the ellipses denote contributions with higher powers of the fields.

Thus it is convenient to consider the combinations
\begin{equation}
G^{(u)} = \left( G^{(1)} + G^{(2)} \right) \quad
G^{(d)} = \left( G^{(1)} - G^{(2)} \right)
\end{equation}
corresponding to the couplings for the neutral currents for the
up-type and the down-type quarks. 

We may now formulate our second assumption: {\it We shall assume that} 
\begin{equation} \label{MFV}
\left[ m^u m^{u\dag}\, , \, G^{(u)} \right] = 0 =
\left[ m^d m^{d\dag}\, , \, G^{(d)} \right],
\end{equation}
{\it which avoids FCNCs at least at tree level.} 

\section{Impact on the charged currents}
The contribution of the operators (\ref{LL1}) and (\ref{LL2})
to the charged current is
\begin{equation} \label{cc}
O^{(1)}_{LL} + O^{(2)}_{LL} = v^2 \left(G_{AB}^{(u)} + G_{AB}^{(d)} \right)
\frac{g}{2 \sqrt{2}} \overline{Q}_A \gamma^\mu 
\left(\tau^+ W_\mu^+ + \tau^- W_\mu^- \right) Q_B + \cdots 
\end{equation}
where the ellipses denote terms with additional fields. 

In order to discuss the implications of (\ref{cc}) it is convenient
to go to the mass eigenbasis. According to our assumption (\ref{MFV})
both $G^{(u)}$ and $G^{(d)}$ are diagonal in this basis, i.e.
\begin{eqnarray} \label{gdefs}
&& S_L^{(u) \dagger} G^{(u)} S_L^{(u)} = {\rm diag} (g_u, \, g_c, \, g_t)
\equiv G_u \\
\nonumber
&& S_L^{(d) \dagger} G^{(d)} S_L^{(d)} = {\rm diag} (g_d, \, g_s, \, g_b)
\equiv G_d
\end{eqnarray}
where $S_L^{(u/d)}$ are the (unitary) transformations to the mass
eigenbasis for the left-handed up/down-type quarks. Inserting this into
(\ref{cc}) we get 
\begin{equation} \label{Veff}
V_{\indi{eff}} = V_{\indi{CKM}} + \frac{v^2}{2 \Lambda^2}
\left( G_u V_{\indi{CKM}} + V_{\indi{CKM}} G_d \right) 
\end{equation}
where
\begin{equation}
V_{\indi{CKM}} = S_L^{(u) \dagger} S_L^{(d)}  
\end{equation}
is the usual definition of the (unitary) CKM matrix from the diagonalisation
of the mass matrices\footnote{This may in fact include the two-point
contributions of ${\mathcal O}^{(1)}_{LR}$, which in this way are completely
absorbed.}.

Due to the fact that both $G^{(u)}$ and $G^{(d)}$ have to be hermitean,
we find that (\ref{gdefs}) defines six real quantities parametrising
possible new-physics effects occuring in the charged current.
Likewise, due to $SU(2) \times U(1)$ symmetry the same parameters
appear in the (flavour-diagonal) neutral currents. 

The implications of (\ref{Veff}) are best analysed by studying the
relations
\begin{eqnarray} \label{unitary1}
V_\indi{eff}^\dagger V_\indi{eff} &=&
1 + \frac{v^2}{\Lambda^2}
    \left(V_\indi{CKM}^\dag  G_u V_{\indi{CKM}} + G_d \right)
\\\label{unitary2}
 V_\indi{eff} V_\indi{eff}^\dagger &=& 1 + \frac{v^2}{\Lambda^2}
    \left(G_u + V_\indi{CKM} G_d V_{\indi{CKM}}^\dag \right).
\end{eqnarray}
We note that $V_\indi{eff}$ is unitary, if 
\begin{equation} \label{g6}
G_u V_{\indi{CKM}} = - V_\indi{CKM} G_d 
\end{equation}
which is equivalent to
\begin{equation}
\frac{v^2}{\Lambda^2} G_u = - \frac{v^2}{\Lambda^2} G_d = g_0 \mathbbm{1}
\end{equation}
where $g_0$ is a real parameter. 
Although the charged currents will then be as in the standard model,
the neutral currents are still affected, see (\ref{nc}).

Similarly, if both $G_u$ and $G_d$ are proportional to the unit matrix
\begin{equation}
\frac{v^2}{\Lambda^2} G_u = \left(g_0 + \frac{1}{2} \Delta g \right) \mathbbm{1}
\quad \mbox{ and } \quad  
\frac{v^2}{\Lambda^2} G_d = - \left(g_0 - \frac{1}{2} \Delta g \right) \mathbbm{1} 
\end{equation}
with another real parameter $\Delta g$ we find that the effective
CKM matrix is proportional to $V_\indi{CKM}$
\begin{equation}
V_\indi{eff} = \left(1 + \frac{1}{2} \Delta g\right)\,  V_\indi{CKM}  \, .
\end{equation}

The elements $V_\indi{eff,ud}$ and $V_\indi{eff,us}$ of the first row of the unitarity
triangle have been measured quite precisely. For this first row we have
\begin{eqnarray}
|V_\indi{eff,ud}|^2 + |V_\indi{eff,us}|^2 + {\cal O} (\lambda^6) 
&=& 1 + \Delta g +
g_1^{(u)} + |V_\indi{eff,ud}|^2 g_1^{(d)}  + {\cal O} (\frac{v^4}{\Lambda^4})
\end{eqnarray}
where we here and in the following use the notation
\begin{equation}
  \begin{array}{>{\displaystyle}rc>{\displaystyle}l@{\qquad\ }>{\displaystyle}rc>{\displaystyle}l}
   g_1^{(u)} &=&\, \frac{v^2}{\Lambda^2}\left(g_u - g_c \right) &
g_2^{(u)}  &=& \frac{v^2}{\Lambda^2}\left(g_t - g_c \right) \\
g_1^{(d)} &=&\, \frac{v^2}{\Lambda^2}\left(g_d - g_s \right) &
g_2^{(d)} & =& \frac{v^2}{\Lambda^2}\left(g_b - g_s \right) \\
g_0\  &=&  \frac{v^2}{2 \Lambda^2} \left(g_c - g_s\right)&
\Delta g &=&  \frac{v^2}{\Lambda^2} \left(g_c + g_s\right)
  \end{array}
\end{equation}
which defines the six parameters of our parametrisation.

Currently there is a statistically insignificant deviation from CKM unitarity
\begin{equation}
|V_\indi{eff,ud}|^2 + |V_\indi{eff,us}|^2 = 0.9957 \pm 0.0026 \, ; 
\end{equation}
speculating that this deviation is due to $G_u$ and $G_d$, we get
for the parameters
\begin{equation}
 \Delta g +
g_1^{(u)} + |V_\indi{eff,ud}|^2 g_1^{(d)} = 0.0043 \pm 0.0026 \,.
\end{equation}

Most of the other tests of the flavour sector of the standard model
usually involve
the off-diagonal elements of (\ref{unitary1}) and (\ref{unitary2}). 
In order to obtain non-diagonal contributions on the right-hand sides
of relations (\ref{unitary1}) and (\ref{unitary2}) $G_u$ and/or $G_d$
have to have different
eigenvalues. Looking at the non-diagonal elements of (\ref{unitary1})
and (\ref{unitary2})
we get
\begin{eqnarray}
\left(V_\indi{eff}^\dagger V_\indi{eff}\right)_{d_id_j} &=&
\frac{v^2}{\Lambda^2}\left(
V^*_{ud_i} V_{ud_j} g_u +
V^*_{cd_i} V_{cd_j} g_c +
V^*_{td_i} V_{td_j} g_t  \right)\quad\ i\not=j\\
\left(V_\indi{eff} V_\indi{eff}^\dagger\right)_{u_iu_j} &= &
\frac{v^2}{\Lambda^2}\left(
V_{u_id} V^*_{u_jd} g_d +
V_{u_is} V^*_{u_js} g_s +
V_{u_ib} V^*_{u_jb} g_b \right)\quad i\not=j
\end{eqnarray}
where $V_{u_i \, d_j}$ are the elements of $V_\indi{CKM}$. 

Making use of the unitarity relation for $V_\indi{CKM}$ we get
\begin{eqnarray} \label{UTd}
\left(V_\indi{eff}^\dagger V_\indi{eff}\right)_{ds} &= & g_1^{(u)}\,
V^*_{ud} V_{us} + g_2^{(u)} \, V^*_{td} V_{ts}\label{UTds}\\
\left(V_\indi{eff}^\dagger V_\indi{eff}\right)_{db} &= & g_1^{(u)}\,
V^*_{ud} V_{ub} + g_2^{(u)}\,  V^*_{td} V_{tb}\label{UTB}\label{UTdb}\\
\left(V_\indi{eff}^\dagger V_\indi{eff}\right)_{sb} &= & g_1^{(u)}\,
V^*_{us} V_{ub} + g_2^{(u)}\,  V^*_{ts} V_{tb}\label{UTsb}\\
\left(V_\indi{eff} V_\indi{eff}^\dagger\,\right)_{uc} &= & 
g_1^{(d)}\,  V_{ud} V^*_{cd} + g_2^{(d)}  V_{ub} V^*_{cb} \label{UTuc}\\
\left(V_\indi{eff} V_\indi{eff}^\dagger\right)_{ut} &= & 
g_1^{(d)} \, V_{ud} V^*_{td} + g_2^{(d)}\,  V_{ub} V^*_{tb}\label{UTut} \\
\left(V_\indi{eff} V_\indi{eff}^\dagger\right)_{ct} &= & 
g_1^{(d)} \, V_{cd} V^*_{td} + g_2^{(d)} \, V_{cb} V^*_{tb} \label{UTct}\label{UTu}
\end{eqnarray}

\begin{figure}[htbp]
  \begin{center}
    \includegraphics[width=.75\textwidth]{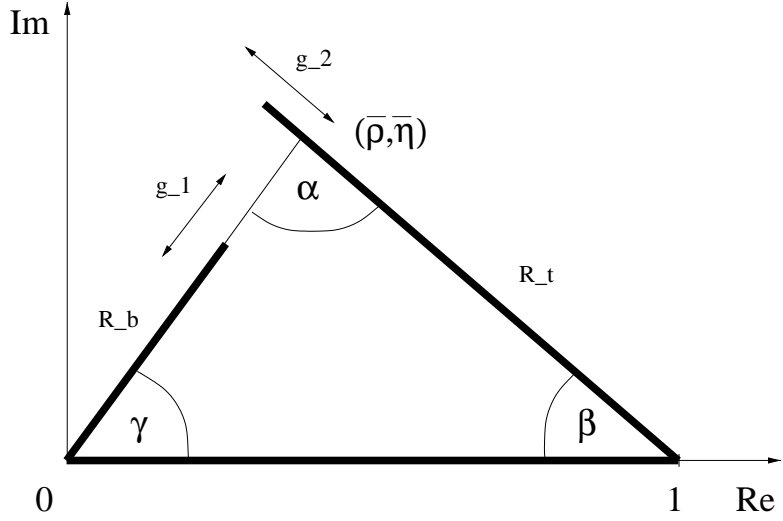}
    \caption{Unitarity ``triangle'' for $V_\indi{eff}$}
    \label{myckm}
  \end{center}
\end{figure}

In the SM the relations (\ref{UTd}) to (\ref{UTu}) have vanishing
left-hand sides and are usually
visualised as triangles in the complex plane. Here the left-hand sides
are not vanishing which corresponds to ``open triangles''. For the $B$~physics triangle the situation is depicted in figure~\ref{myckm}.

An expansion in powers of the Wolfenstein parameter $\lambda$ shows 
that in (\ref{UTds}) the $g^{(u)}_2$ contribution is suppressed
by a factor of $\lambda^4$ compared to the $g^{(u)}_1$ part. Therefore
a measurement of this triangle is sensive to new-physics effects coming
from $g^{(u)}_1$. 
We get with data from \cite{PDG}
\begin{equation}
  \left|1-g_1^{(u)}\right|  + \mathcal{O}(\lambda^4)=
  \left|\frac{V_\indi{eff,cd}^*V_\indi{eff,cs}}{V_\indi{eff,ud}^*V_\indi{eff,us}}\right| 
= 1.044 \pm 0.076
\end{equation}
which translates into
\begin{equation}
g_1^{(u)} =  -0.044 \pm 0.076\,.
\end{equation}

Nearly the same situation comes from the unitarity triangle
(\ref{UTsb}) where the coefficient for $g^{(u)}_1$ is of order
$\lambda^2$ which makes this triangle sensitive to $g^{(u)}_2$
only. Unfortunately we cannot give an upper limit for $|g^{(u)}_2|$ because there is
no precise measurement for $|V_{ts}|$ yet. 

The usual $B$~physics unitarity triangle is given by (\ref{UTB}). Here
the coefficients for the $g^{(u)}_{1/2}$ are of the same order which
makes the measurement of this triangle sensitive to both
coefficients. In the following we will mainly discuss this triangle
since it is
in the main focus of the $B$ factories. For the triangles containing the
coefficients $g^{(d)}_{1/2}$ the discussion is analogous.

Using only matrix elements of $V_\indi{eff}$ we obtain for (\ref{UTB})
\begin{eqnarray}
\left(V_\indi{eff}^\dagger V_\indi{eff} \right)_{bd} &=&
g_1\, V_{\indi{eff,ud}} V_{\indi{eff,ub}}^* + g_2
V_{\indi{eff,td}} V_{\indi{eff,tb}}^*
+ \mathcal{O}(\textstyle{\frac{v^4}{\Lambda^4}})
\end{eqnarray}
where we have dropped the superscript $(u)$, since we shall only
consider this unitarity triangle for the rest of the paper. 
Furthermore, we can formulate all triangle relations with 
$V_\indi{eff}$ only; therefore $V$ means $V_\indi{eff}$ from now on.
In particular, all matrix elements $V_{u_i \, d_j}$ are now elements
of  $V \equiv V_\indi{eff}$ .

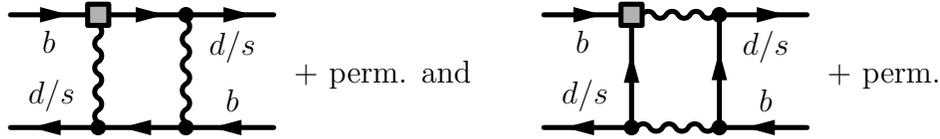
\begin{figure}[htb]
  \begin{center}\mbox{}\\[3mm]
    \begin{fmffile}{HMboxdiag}
\begin{fmfgraph*}(35,15)
  \fmfstraight
  \fmfpen{thick}
  \fmftop{t1,t2,t4,t5}
  \fmfbottom{b1,b2,b4,b5}
  \fmf{fermion,label=$b$}{t1,t2}
  \fmf{fermion,label=$d/s$}{b2,b1}
  \fmf{fermion,label=$b$}{b5,b4}
  \fmf{fermion,label=$d/s$}{t4,t5}
  \fmf{boson}{b2,t2}
  \fmf{boson}{b4,t4}
  \fmf{fermion}{t2,t4}
  \fmf{fermion}{b4,b2}
  \fmfdot{t4,b2,b4}
  \fmfv{decor.shape=square,decor.size=4thick,decor.filled=30}{t2}
  \end{fmfgraph*}
 \raisebox{6mm}{\ + perm. and\qquad}
\begin{fmfgraph*}(35,15)
  \fmfstraight
  \fmfpen{thick}
  \fmftop{t1,t2,t4,t5}
  \fmfbottom{b1,b2,b4,b5}
  \fmf{fermion,label=$b$}{t1,t2}
  \fmf{fermion,label=$d/s$}{b2,b1}
  \fmf{fermion,label=$b$}{b5,b4}
  \fmf{fermion,label=$d/s$}{t4,t5}
  \fmf{fermion}{b2,t2}
  \fmf{fermion}{b4,t4}
  \fmf{boson}{t2,t4}
  \fmf{boson}{b4,b2}
  \fmfdot{t4,b2,b4}
  \fmfv{decor.shape=square,decor.size=4thick,decor.filled=30}{t2}
  \end{fmfgraph*}
 \raisebox{6mm}{\ + perm.}
    \end{fmffile}

\end{center}
  \caption{Diagrams corresponding to the new-physics effects in
  $B$-$\overline{B}$ mixing.}
\label{BBbar}
\end{figure}

In the standard analysis of the $B$~physics unitarity triangle
the ``$c$-side'' serves as the normalised base; the remaining
sides are as usual 
\begin{equation}
R_b = \left| \frac{V_{ud} V^*_{ub}}{V_{cd} V^*_{cb}}\right| \qquad
\mathrm{and}\qquad R_t = \left| \frac{V_{td} V^*_{tb}}{V_{cd} V^*_{cb}}\right|.
\end{equation}
They are given by the measurement of semileptonic $B$ decays and
by the oscillation frequency of $B$-$\overline{B}$ mixing. Note that 
$B$-$\overline{B}$ oscillations are given by a loop process involving
standard model particles. In our approach a modification of $\Delta m_{d/s}$ is
due to the diagrams depicted in fig~\ref{BBbar}, which is due to our
assumptions the only source for
a new effect. This is in contrast to the usual point of view, where non-SM effects
are induced by heavy particles in the loops, which - in the case of
$B$-$\overline{B}$ mixing - leads to a four fermion operator of the form
$(\bar{b} d)(\bar{b} d)$ which has been considered elsewhere \cite{Grossman:1997dd}.

Furthermore, the angles are then given by
\begin{equation}
\alpha = \arg\left(-\frac{V_{td} V^*_{tb}}{V_{ud} V^*_{ub}}\right)
\qquad
\beta  = \arg\left(-\frac{V_{cd} V^*_{cb}}{V_{td} V^*_{tb}}\right)
\qquad
\gamma = \arg\left(-\frac{V_{ud} V^*_{ub}}{V_{cd} V^*_{cb}}\right).
\end{equation}
Expressed in these quantites, the 
unitarity relation may be cast into the form
\begin{equation} \label{UTb}
\left(1-g_1\right) R_b e^{i \gamma} +
\left(1-g_2\right) R_t e^{-i \beta} = 1 .
\end{equation}

A few comments are in order. Since the parameters $g_i$ are real,
there is no additional source for CP violation in the proposed
parametrisation. For this reason the angles appearing in (\ref{UTb})
are the same as in the original $V_\indi{CKM}$. Thus 
the effects of the new physics show up in
the lengths of the sides of the unitarity triangles only, which appear
``stretched'' by the factors $g_i$. The situation is
sketched in figure \ref{myckm}.

It is interesting to note that with the current inputs ($\sin (2 \beta)$,
$|V_{ub}|/|V_{cb}|$ and $\Delta m_{d/s}$) \cite{PDG} we still do not get any
constraint on the parameters $g_i$. The main problem is our ignorance
of the angle $\gamma$. Even if we include the information from Kaon
physics, we obtain a wide range for $\gamma$ 
\begin{equation}
30^\circ \le \gamma \le 165^\circ 
\end{equation}
from the intersections of the hyperbola from
Kaon-CP violation and the circle from $|V_{ub} / V_{cb}|$.
If this angle were known better (e.g. from a measurement
of a CP asymmetry) we could constrain the parameters $g_i$.

\begin{figure}[htbp]
  \begin{center}
    \includegraphics[width=.75\textwidth]{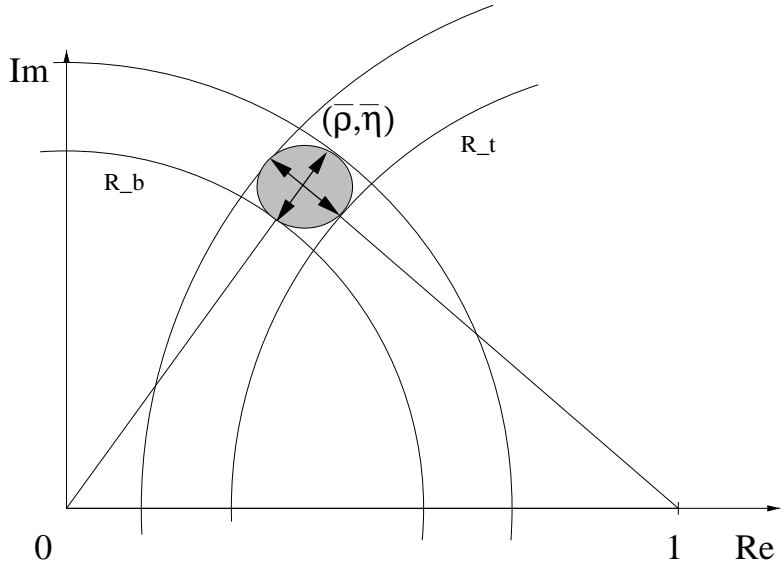}
    \caption{Unitarity ``triangle'', assuming the SM range for $\gamma$.}
    \label{maximal}
  \end{center}
\end{figure}

A measurement of $\gamma$ is currently not available, so we shall
proceed by making an assumption for the angle $\gamma$. We shall
{\em assume} that a measurement of $\gamma$ as been performed which
yields $\gamma$ in the range of the current standard-model fits. 
The possible range for $g_1$ and $g_2$ is sketched in fig.~\ref{maximal}.
Taking the current $95\%$ confidence levels of the CKM-Fit
\cite{CKMFIT} we find for the possible ranges 
\begin{equation}
 -0.49 \le g_1 \le +0.33\qquad 
 -0.54 \le g_2 \le +0.35 
\end{equation}
The $g_i$ are of order $\frac{v^2}{\Lambda^2}$ which translates
into a  limit of roughly $\Lambda \ge 500\, {\rm GeV}$

\section{Implications for neutral currents}
Finally we have to consider the effect on neutral currents, in particular on
flavour changing neutral currents (FCNCs). The tree-level contributions to
FCNCs are absent by construction, but through loops the charged currents
induce modifications of the neutral currents. The insertion of the effective
CKM  matrix for the charged currents induces a violation of the GIM mechanism
which leads to FCNCs already at one loop. 

\begin{figure}[htb]
  \begin{center}\mbox{}\\[3mm]
    \begin{fmffile}{HMZoneloop}
\vspace{1mm}
\begin{fmfgraph*}(50,20)
  \fmfpen{thick}
  \fmfleft{a0,a1,i,a2}
  \fmfright{b0,b1,o,b2}
  \fmfbottom{z}
  \fmf{fermion,label=$d$,label.side=left}{i,v1}
  \fmf{phantom,tension=2}{v1,v2,v3}
  \fmf{fermion,label=$d'$,label.side=left}{v3,o}
  \fmfv{decor.shape=square,decor.size=5thick,decor.filled=30}{v1}
  \fmfdot{v3}
  \fmffreeze
  \fmf{boson,label=$W$,label.side=right,left}{v1,v3}
  \fmf{boson,label=$Z$,label.side=left}{v2,z}
  \fmf{fermion}{v1,v2,v3}
  \end{fmfgraph*}
\qquad
\begin{fmfgraph*}(50,20)
  \fmfpen{thick}
  \fmfleft{a0,a1,i,a2}
  \fmfright{b0,b1,o,b2}
  \fmfbottom{z}
  \fmf{fermion,label=$d$,label.side=left}{i,v1}
  \fmf{phantom,tension=2}{v1,v2,v3}
  \fmf{fermion,label=$d'$,label.side=left}{v3,o}
  \fmfv{decor.shape=square,decor.size=5thick,decor.filled=30}{v1}
  \fmfdot{v3}
  \fmffreeze
  \fmf{fermion}{v1,v3}
  \fmf{boson,label=$W$,label.side=left,right,tag=1}{v1,v3}
  \fmf{phantom,tag=2}{v2,z}
  \fmfipath{p[]}
  \fmfiset{p1}{vpath1(__v1, __v3)}
  \fmfiset{p2}{vpath2(__v2, __z)}
  \fmfi{boson,label=$Z$,label.side=left}{point length(p1)/2 of p1 -- point length(p2) of p2}      
  \fmfiv{decor.shape=circle,decor.filled=full,decor.size=2thick}{point length(p1)/2 of p1}
\end{fmfgraph*}
\end{fmffile}
\end{center}
  \caption{Diagrams leading to FCNCs on the one-loop level. This
  corresponds to a mixing of the Operators $O_{LL}$.}
\label{Zoneloop}
\end{figure}
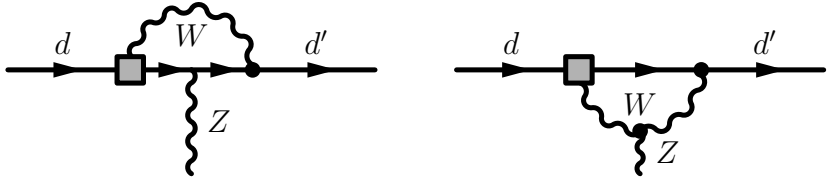

The obvious example for such an effect is an effective left-handed FCNC
coupling to the $Z$ boson induced by the loop diagrams shown in
fig.~\ref{Zoneloop}. They lead to a mixing of the operators 
$O_{LL}$ such that off-diagonal contributions to $O^{(2)}_{LL}$ and
the neutral components of  $O^{(1)}_{LL}$ appear.
These contributions are expected to be small, although
they are enhanced by a large logarithm of the type $\ln (\Lambda / M_W)$
where $\Lambda$ is the scale of new physics. The suppression originates
on the one hand from the electroweak loop factor $g^2 /(16 \pi^2)$ and
on the other hand from the fact that this contribution is proportional to
the GIM violation (\ref{unitary1}, \ref{unitary2}), which has to be a small
quantity as well.  

Another effect of similar type is the modification of the 
$\rho$ parameter, whose deviation from
unity is defined as usual by 
\begin{equation}
\Delta \rho = \frac{A_{ZZ} (0)}{M_Z^2} - \frac{A_{WW} (0)}{M_W^2}
\end{equation}
where $A_{ZZ}$ and $A_{WW}$ are the transverse contributions to the
$Z$ and the $W$ self energies.

Clearly there is a contribution to the $\rho$ parameter from
dimension-six operators which contain only Higgs and gauge fields \cite{Georgi}.
An example is the operator
\begin{equation} \label{drhoop}
R = {\rm Tr} \left\{ L_\mu  L_3^\mu \right\} 
\end{equation}
which leads at tree level to a modification of $\rho$. 

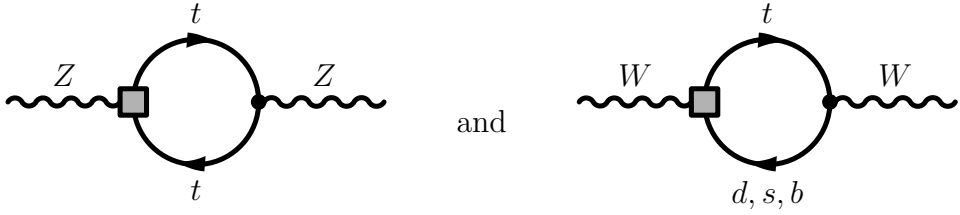
\begin{figure}[htb]
  \begin{center}\mbox{}\\[3mm]
    \begin{fmffile}{HMrhofigure}
\begin{fmfgraph*}(50,20)
  \fmfpen{thick}
  \fmfleft{i}
  \fmfright{o}
  \fmf{boson,label=$Z$,label.side=left,tension=2}{i,v1}
  \fmf{boson,label=$Z$,label.side=left,tension=2}{v2,o}
  \fmf{fermion,left,label=$t$,label.side=left}{v1,v2}
  \fmf{fermion,left,label=$t$,label.side=left}{v2,v1}  
  \fmfv{decor.shape=square,decor.size=5thick,decor.filled=30}{v1}
  \fmfdot{v2}
  \end{fmfgraph*}
\raisebox{6mm}{\qquad and\qquad}
\begin{fmfgraph*}(50,20)
  \fmfpen{thick}
  \fmfleft{i}
  \fmfright{o}
  \fmf{boson,label=$W$,label.side=left,tension=2}{i,v1}
  \fmf{boson,label=$W$,label.side=left,tension=2}{v2,o}
  \fmf{fermion,left,label=$t$,label.side=left}{v1,v2}
  \fmf{fermion,left,label=$d,,s,,b$,label.side=left}{v2,v1}
  \fmfv{decor.shape=square,decor.size=5thick,decor.filled=30}{v1}
  \fmfdot{v2}
  \end{fmfgraph*}
\end{fmffile}
\end{center}
  \caption{One-loop diagrams for the new-physics contribution to
           the $\rho$ parameter. The box indicates the insertion of a
           vertex corresponding to $O_{LL}^{(1)}$ and $O_{LL}^{(2)}$.}
\label{fig2}
\end{figure}

However, in our case we can 
consider the contribution of (\ref{LL1}) and (\ref{LL2}) to the
$\rho$ parameter by computing the diagrams shown in fig.~\ref{fig2}.

In the standard model the $\rho$ parameter is convergent due to the
fact that the divergencies of the charged and the neutral current cancel
exactly. Including the new-physics contributions disturbs this cancellation,
leading to an enhancement by a logarithm of the
scale of the new-physics contribution. This indicates that a mixing
of the operators (\ref{LL1}) and (\ref{LL2}) into operators of the type 
(\ref{drhoop}) occurs. 

Keeping only the dominant
contribution from the top quark we obtain 
\begin{equation} \label{rhonew}
\Delta \rho = \frac{3 G_F m_t^2}{2 \pi^2 \sqrt{2}}
\left(-2 g_0 - g_2^{(u)} + g_1^{(d)} |V_{td}|^2 +  g_2^{(d)} |V_{tb}|^2 \right)
\ln \left(\frac{\Lambda^2}{M_W^2} \right).
\end{equation}

We note that the operator (\ref{LL1}) conserves the custodial $SU(2)_C$,
while (\ref{LL2}) breaks this symmetry. The contribution of (\ref{LL2})
is proportional to the difference $v^2/\Lambda^2 \left(G_u - G_d\right) = 2 g_0 \mathbbm{1}$, while
the $SU(2)_C$ conserving piece is $v^2/\Lambda^2 \left(G_u + G_d\right) = \Delta g \mathbbm{1}$.
The $\rho$-parameter is a measure of $SU(2)_C$ breaking and hence it
cannot depend on $\Delta g$. Although the case
$v^2/\Lambda^2\, G_u = - v^2/\Lambda^2\, G_d = g_0 \mathbbm{1}$ corresponds to the case where the
CKM matrix is unitary despite a possible new-physics contribution,
it still changes the strength of the charged current relative to the
neutral one, and thus this appears in the $\rho$ parameter. In turn
$v^2/\Lambda^2 \left(G_u + G_d\right) = \Delta g \mathbbm{1}$ changes both the coupling of the
neutral as well as the charged currents by the same amount, but this
has to lead to a non-unitary CKM matrix with
$V_\indi{eff} V_\indi{eff}^\dag = (1+ \Delta g)\mathbbm{1}$.

Likewise, the remaining terms in (\ref{rhonew}) have a similarly
simple explanation. Already the mass matrices of the standard model
violate $SU(2)_C$ leading to nontrivial mixings. If this effect
was absent, we would have $V_{td} = 0$ and $V_{tb} = 1$. If now the new
physics effects would conserve $SU(2)_C$ we would have
$g_2^{(u)} = - g_2^{(d)}$ in which case the $\rho$ parameter again
would not be affected.  

Finally we may also consider the effect of the new-physics
contributions in (\ref{LL1}) and (\ref{LL2}) on the forward-backward
asymmetry for bottom quarks produced in $e^+ e^-$ collisions on the $Z$
resonance. On resonance the forward-backward asymmetry is given by 
\begin{equation}
A_{FB} = \frac{3}{4}
\left(\frac{g_{L,e}^2 - g_{R,e}^2}{g_{L,e}^2 + g_{R,e}^2}\right)
\left(\frac{g_{L,b}^2 - g_{R,b}^2}{g_{L,b}^2 + g_{R,b}^2}\right)
\end{equation}
where $g_{L,i}$ and $g_{R,i}$ are the left- and right-handed couplings of
the particle $i$.

According to (\ref{LL1}) and (\ref{LL2}) we assume that only the
left-handed couplings of the bottom quark deviate from the standard model
values. Inserting the couplings of the electron and the bottom quark
we get
\begin{eqnarray}
A_{FB} &=& \frac{9 (4 s_W^2 - 3)(4 s_W^2 - 1)}
              {4(8 s_W^4 -12 s_W^2 + 9)(8 s_W^4 -4 s_W^2 + 1)}
\\ \nonumber
&& \quad - \frac{36 s_W^4 (2 s_W^2 -3)(4 s_W^2 -1) }
              {(8 s_W^4 -12 s_W^2 + 9)^2(8 s_W^4 -4 s_W^2 + 1)} \, g_b 
\end{eqnarray}
where $ s_W^2 = \sin^2 \Theta_W$ is the weak mixing parameter and $g_b$ is
of order $v^2 / \Lambda^2$.

It is interesting to note that the coefficient in front of the new-physics
parameter $g_b$ is very small; putting in $ s_W^2 \approx  0.2314$
which reproduces the best fit for the SM \cite{renton} one obtains
\begin{equation}
A_{FB} = 0.1039 - 0.016 g_b
\end{equation}
which makes such an effect hard to observe. Currently there is a statistically
insignificant deviation of the measured value $A_{FB} = 0.0994 \pm 0.0017$
from the standard model expectation. Attributing this to $g_b$ yields
$g_b \sim 0.3$ which is quite enormous, because $g_b$ is of order
$v^2 / \Lambda^2$ and such a value would lead to a relatively low $\Lambda$ of order
500 GeV.

\section{Conclusions}
We suggest a possible parametrisation of new-physics effects
in flavour physics. It is based on considerations of dimension-six
operators, out of which we have discussed the operators with 
two quark fields only. Making two more assumptions which is flavour 
conservation and the existence of a massless limit we can reduce
the number of new-physics parameters to six.

We discussed the impact of these contributions on charged as
well as on neutral currents. In the sector of charged currents, our
parametrisation affects the analysis of the $B$ physics unitarity
triangle in a well defined way by replacing the CKM matrix by an
effective one. However, with present data most of the parameters
cannot be constrained significantly. 
Due to the fact that the effective CKM matrix is not necessarily
unitary anymore flavour changing neutral currents receive
contributions at one loop from the violation of the GIM mechanism
which we assume to be small. Furthermore, in the neutral
currents the $SU(2)_C$ violating parameters affect the $\rho$ parameter
yielding a possible contribution of order
$M_W^2 / \Lambda^2 \ln (\Lambda^2 / M_W^2)$; however, the neutral
currents test a different set of parameters as the CKM analysis. Finally, the
forward backward asymmetry in $e^+ e^- \to Z_0 \to \bar{b} b$ is not
very sensitive due to a small coefficient.

Clearly the analysis of the above set of operators cannot cover the full
variety of possible new-physics contributions in the flavour sector. 
In particular, the set of all possible four-fermion operators 
has a nontrivial flavour structure, but is very large. Consequently
one has to impose additional assumptions concerning these operators to
deal with them in practical applications.

One possibility, which has been discussed already in \cite{BW}, is to
impose additional symmetries such as a horizontal or family symmetry.
However, in general the couplings of the charged and neutral currents
turn out to be of similar size. Using the stringent constraints on FCNCs
yields very small couplings for the neutral current operators, which
in turn then implies that also the charged current contributions will
have very small couplings. At least in scenarios of this type it is
justified to neglect the four-fermion operators e.g. in the analysis
of the $B$ physics unitarity triangle discussed above.

\section*{Acknowledgement}

The authors thank A.~Buras, A.~Falk, G.~Isidori, W.~Kilian and U.~Nierste for discussions. 
TM and TH acknowledge the support of the DFG Sonderforschungsbereich
SFB/TR9 ``Computational Particle Physics'' and the DFG Graduiertenkolleg
``High Energy Physics and Particle Astrophysics''. TH is also supported
by a fellowship of the ``Studienstiftung des Deutschen Volkes''.

\end{document}